\documentclass[sigconf]{acmart}

\usepackage{xspace}
\usepackage{amsmath}

\usepackage{multirow}
\usepackage{subcaption}
\usepackage{graphicx}	
\usepackage{booktabs}
\usepackage{xcolor}
\usepackage{paralist}
\usepackage{enumitem,kantlipsum}
\usepackage{balance}
\usepackage{colortbl}

\AtBeginDocument{%
  }

\begin{document}
\title[]{The Importance of Cognitive Biases in the Recommendation Ecosystem}


\author{Markus Schedl$^{1,2}$, Oleg Lesota$^{1,2}$, Stefan Brandl$^{2}$, Mohammad Lotfi$^{2}$, Shahed Masoudian$^{1}$}
\email{markus.schedl@jku.at}
\orcid{0000-0003-1706-3406}
\affiliation{
  \institution{$^{1}$ Johannes Kepler University Linz and \\ $^{2}$ Linz Institute of Technology}
  \city{Linz}
  \country{Austria}
}

\renewcommand{\shortauthors}{Schedl}

\settopmatter{printacmref=false}
\renewcommand\footnotetextcopyrightpermission[1]{}
\pagestyle{plain}
\begin{abstract}
Cognitive biases have been studied in psychology, sociology, and behavioral economics for decades. Traditionally, they have been considered a negative human trait that leads to inferior decision-making, reinforcement of stereotypes, or can be exploited to manipulate consumers, respectively.
We argue that cognitive biases also manifest in different parts of the recommendation ecosystem and at different stages of the recommendation process. More importantly, we contest this traditional detrimental perspective on cognitive biases and claim that certain cognitive biases can be beneficial when accounted for by recommender systems.
Concretely, we provide empirical evidence that biases such as feature-positive effect, Ikea effect, and cultural homophily can be observed in various components of the recommendation pipeline, including input data (such as ratings or side information), recommendation algorithm or model (and consequently recommended items), and user interactions with the system. 
In three small experiments covering recruitment and entertainment domains, we study the pervasiveness of the aforementioned biases. We ultimately advocate for a prejudice-free consideration of cognitive biases to improve user and item models as well as recommendation algorithms. 
\end{abstract}



\maketitle

\section{Introduction and Background}
\textit{``Music used to be better in the 1980s when I was young.''}\\ 
\textit{``The cookies I baked are much tastier than the ones I bought.''}\\
\textit{''I only remember the last items on my to-do list.''}\\
\textit{``He is such a great actor; I am sure those nasty accusations against him are made up.''}

These are examples of common cognitive biases, respectively, declinism, Ikea effect, recency effect, and halo effect.
Such biases have been studied in psychology and sociology 
for decades. In psychology, they are commonly defined as systematic perceptual deviations of the individual from rationality and objectivity, cognition, judgment, or decision-making, which often happens unconsciously~\cite{daniel2017thinking,dobelli2013art}.
In sociology, they typically refer to collective prejudices of a society that favor one group's values, norms, and traditions over others~\cite{eberhardt2020biased,bonilla2006racism}.

Historically, cognitive biases have been regarded as a negative characteristic of humans, which lead to inferior decision-making, reinforce stereotypes, and may even cause severe systematic errors and harm~\cite{tversky1974judgment}. 
Only recently, psychologists have started to acknowledge the positive effects of certain cognitive biases, e.g., to improve the efficiency of human learning and decision-making~\cite{gigerenzer2011heuristic,daniel2017thinking}.
In the field of machine learning, the study of cognitive biases has played a very minor role so far. Only recently, some ideas to leverage cognitive biases for model training, e.g., to improve their generalization capabilities or foster ethical machine behavior emerged~\cite{DBLP:journals/corr/abs-2203-09911}.
Narrowing down the scope to search and ranking, in information retrieval, some preliminary research on cognitive biases has been conducted recently~\cite{azzopardi2021cobisearch,10.1145/3626772.3661382,DBLP:journals/jis/GomrokiBFF23}.
Existing research, however, has focused on detecting some cognitive biases and assessing their influence on search behavior rather than leveraging them to improve retrieval algorithms.

In recommender systems (RSs) research, some psychologically grounded human biases have been studied in the past, e.g., primacy and recency effects in peer recommendation~\cite{10.1371/journal.pone.0098914} as well as risk aversion and decision biases in product recommendation~\cite{10.1080/15332861.2015.1018703}. However, cognitive biases in the context of recommendation have barely been studied lately. Nor are we aware of any systematic investigation of their manifestations at different stages of the recommendation process. This is particularly astonishing because research in RSs has historically been 
inspired by psychological theories, models, and empirical evidence on human decision-making~\cite{DBLP:journals/ftir/LexKSTFS21}.


To narrow this research gap, we propose a \textit{holistic examination of cognitive biases within the recommendation ecosystem}. And we take first tiny steps in this direction in the paper at hand.
We advocate for studying their potential manifestations in different \textit{components} of the system, at different \textit{stages} of the recommendation process, and from the perspective of different \textit{stakeholders}. 
Furthermore, we aim to evaluate and harness the positive effects these biases may have, with the goal of enhancing user and item models, as well as refining recommendation algorithms.

In the following, we briefly introduce a selection of cognitive biases that we believe deserve a more thorough investigation in the context of RSs (Section~\ref{sec:cobis}). We then present formalizations and empirical evidence of some of these biases and showcase how they may influence recommendation (Section~\ref{sec:findings}).
Ultimately, we argue for a much-needed (re-)consideration of cognitive biases and point to important directions this could take (Section~\ref{sec:conclusions}).



\section{Cognitive Biases in Recommendation}\label{sec:cobis}
Extensive research in psychology, sociology, and economics has revealed a plethora of cognitive biases~\cite{daniel2017thinking,dobelli2013art}. They relate to how humans perceive, process, store, and retrieve information, involving the cognitive and neurological processes of individuals and even whole societies.
While not all cognitive biases directly apply to RSs, many of them influence user behavior and decision-making processes. As a result, these biases affect users' interactions with items (e.g., ratings or consumption patterns) and with RSs in general (e.g., use of different functionalities provided by the system's interface). However, only a few cognitive biases have been studied in the context of RSs. 
In the following, we introduce some, point to related work, and provide a working definition in the context of the recommendation ecosystem.\footnote{This is, by no means, meant to be an exhaustive list, but a biased personal selection of the authors. We also exclude biases that have already received notable attention in the context of RSs, e.g., confirmation bias and selection bias resulting in interactions or ratings missing-not-at-random.}

\textbf{Feature-Positive Effect \cite{allison1988feature,rassin2023individual}:}
Humans better realize and put more emphasis on things that are present than on things that are absent. 
In the context of RSs, we argue that this effect plays a crucial role in explainability and fairness~\cite{ekstrand2022fairness}. For instance, through counterfactuals, one could show users which (maybe better-suited) items would have been recommended if they had different traits.
We demonstrate this effect in  Section~\ref{sec:experiments:feature_positive}.

\textbf{Ikea Effect \cite{norton2012Ikeaeffect,marsh2018laborleadtolove}:} 
The more effort a person invested in something, the more they will value it.
This effect owes its name to a 
study that found participants were willing to pay a much higher price for furniture they had helped build than for ready-made items~\cite{norton2012Ikeaeffect}. It can be thought of as the desire of humans to justify their efforts.
In an RS context, we assume the existence of similar effects when users create their own item collections (e.g., songs or videos in a playlist, visited places) in contrast to interacting with those created by others or by a recommendation algorithm.
We provide evidence for this effect in Section~\ref{sec:experiments:ikea}.

\textbf{Homophily \cite{homophily2017,mark2003culture}:}
Homophily refers to the fact that individuals tend to form connections with others who have similar characteristics (e.g., age, culture, or religion) more often than with people having different traits.
While it is a well-studied phenomenon in sociology~\cite{mark2003culture} and even evolutionary biology~\cite{jiang2013assortative}, in RSs, homophily has not been extensively studied from a cognitive psychological perspective.\footnote{This is even more surprising because collaborative filtering algorithms leverage this concept, by exploiting similarity in preferences or interaction behaviors.}
Evidence of cultural homophily in the music domain has been found in \cite{DBLP:conf/ismir/LesotaPBLRS22}, where music listeners of certain countries (e.g., Brazil, Sweden, and Germany) showed a preference to listen to domestic music artists. Vice versa, Brazilian, Polish, and Russian artists were found to be predominantly liked by their domestic audience. 
In Section~\ref{sec:experiments:homophily}, we provide additional evidence that such effects can be observed and influenced by the recommendation algorithm. 

\textbf{Declinism \cite{MATHER2005496,doi:10.1177/0956797612459658}:}
The perception that the world or society is declining, i.e.,~things get worse over time. This has been shown to be (partly) the result of \textit{rosy retrospection} --- humans' tendency to remember the past as more positive as it actually was~\cite{mitchell1997temporal}.
Declinism bias can be studied in the context of RSs, by considering side information on items, e.g., emotions reflected in song lyrics have become more negative over the last decades~\cite{parada2024song}.
We hypothesize that such trends can also be observed in longer-term historic interaction data, i.e., users tend to interact more frequently over time with items that are negatively emotion-laden. Identifying and formalizing such trends could also be used to adjust recommendations to counteract (or amplify) this bias.

\textbf{Contrast Effect \cite{10.1177/1541931213601044}:} 
If two items 
are shown in the vicinity in the recommendation list of a user --- one with exceptional, the other with medium utility --- the latter will appear much less appealing to the user, even if it is still a reasonable choice.

\textbf{Anchoring \cite{chapman2002incorporating}:} 
As a variant of the contrast effect, anchoring refers to the fact that humans often overemphasize the piece of information they are exposed to first. 
RS providers can exploit this effect to offer a hook to the user, e.g., showing them a highly-priced item first, making them (unconsciously) believe that subsequent items are cheap even if they are still overpriced~\cite{payne1993adaptive}.
This use is also referred to as \textit{decoy effect}~\cite{teppan2009calculating}.

\textbf{Conformity Bias \cite{adomavicius2011recommender}:}
Providing users with evidence of previous interactions or ratings by others impacts their own consequent behavior.
For instance, showing users (artificial) ratings before asking them to provide their own results in their ratings being closer to the initially presented ones~\cite{adomavicius2011recommender}. 
Also, users are more likely to click on an item if they see that many others have done so~\cite{DBLP:conf/www/ZhengGLHLJ21}.

\textbf{Primacy and Recency Effect (Position Bias) \cite{DBLP:conf/wsdm/CraswellZTR08,10.1371/journal.pone.0098914}:} 
The position at which items occur in a recommendation list influences the probability that users will interact with them.
Users are more likely to interact with items appearing at the beginning (\textit{primacy effect}) and at the end (\textit{recency effect}) of a ranking or sequence of recommendations.
In some specific domains, \textit{position bias} (or \textit{serial position effects}) have been observed already, e.g., users were more likely to vote for recommended science stories shown to them first and last, though the primacy effect was much stronger than the recency effect in this case~\cite{10.1371/journal.pone.0098914}.
In product recommendation, an influence of item positions on users remembering product characteristics and selecting products was observed~\cite{DBLP:conf/persuasive/FelfernigFGHKLMRSTV07}.
Recency effects have also been formalized in the cognitive architecture ACT-R~\cite{anderson2004integrated}, which has been used to predict repeated consumption behavior~\cite{10.1145/3460231.3478846} and to increase diversity and explainability of RSs~\cite{10.1145/3604915.3608838}.

\textbf{Halo Effect \cite{nisbett1977halo,thorndike1920constant}:}
One's overall impression or specific perceived traits of a person influence the perception of other characteristics of that person. For instance, it has been shown that physically attractive people tend to be perceived as more intelligent and having more positive personality traits than less attractive ones~\cite{batres2023examining}. This overly positive perception overshadows the negative traits of the person.
In the context of RSs, we assume that similar effects occur, e.g., in slate-, playlist-, or basket recommendation, where the recommended collection of items as a whole may be perceived as more or less favorable depending on a single trait of one salient item (or its creator) in the recommended item collection.
If evidence for this can be found, RS providers could, for instance, implement a mechanism to push content created by underrepresented producers by adding it to a collection with 
items whose halo will extend to those injected items, thereby improving the latter's exposure and overall fairness.




It is important to note that cognitive biases usually do not occur in isolation. Instead, in the recommendation process, they are often intertwined and related through feedback loops.
For instance, position bias influences the probability of users clicking on a recommended item, which affects the interaction data used to train or update the recommendation model. Also, primacy-, contrast-, and anchoring effects are often at work jointly.

\section{Empirical Findings}\label{sec:findings}
To showcase some manifestations of cognitive biases in the context of RSs, we present several preliminary investigations of a selection of cognitive biases, which are either reflected in the training data, the algorithm outcomes, or user interactions with the system. 
Our experiments cover the domains of recruitment (candidate recommendation) and entertainment (music recommendation). 








\subsection{Feature-Positive Effect}\label{sec:experiments:feature_positive}
The feature positive effect (FPE) is known as the tendency of learning organisms to better detect the presence of stimuli (e.g., \textit{if p then~q}) rather than their absence (e.g., \textit{if p then not~q} or \textit{if not p then~q})~\cite{rassin2023individual,jenkins1969development}. This concept can be relevant in job and candidate RSs, as these systems also tend to focus on what is present in a job ad to determine applicant relevance, potentially overlooking what is missing in the job ad.

To investigate the presence of FPE in job RSs, we evaluate 
a job RS dataset containing 272 job ads and 336 unique applicants across 6 job categories. We predict the match between pairs of candidates and job ads, using a trained RS model following the work of Kumar et al.~\cite{Kumar2023Red_Words_RecSysHR}, and counted 13,607 true positive (TP) and 1,625 false negative (FN) predictions, where TP means the candidate is suitable for the job ad and the model predicts it correctly, while FN means the candidate is suited for the job ad but the model predicts unsuitable. 

\begin{figure} 
    \centering
    \scalebox{0.5}{
    \includegraphics[width=1.25\linewidth]{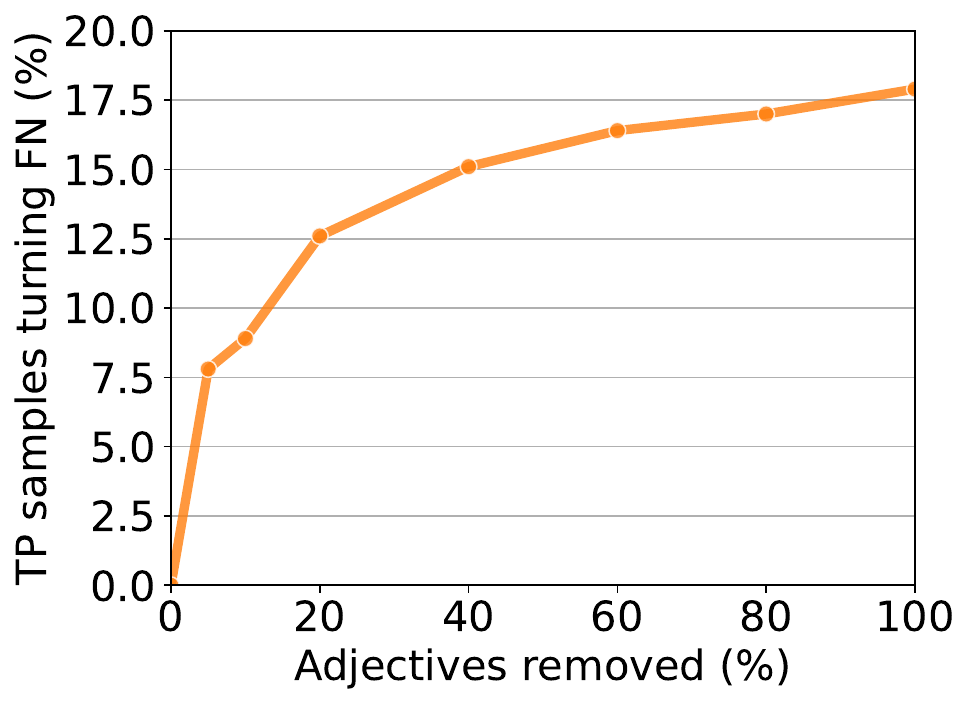}}
    \caption{The effect of removing adjectives from job ads on the relevance predictions of candidate-job pairs.}
    \label{fig:FPE}
\end{figure}

In one experiment, we removed adjectives from TP samples. As shown in Figure~\ref{fig:FPE}, we found that as more adjectives were removed, more TP samples became FNs. This points to the crucial role of adjectives in RSs decision-making of applicants' suitability for a job even though ``a passionate dentist'' should objectively be treated the same as ``a dentist'' when the model evaluates the candidates.

In another experiment, we explored leveraging adjectives by calculating recall for each job ad and grouping them into low- and high-recall categories. We then identified a unique set of adjectives present in high-recall ads but absent in low-recall ads. Please refer to Table~\ref{tab:fpe_adjectives} for some examples. By randomly replacing adjectives in the FN samples with these unique adjectives, we observed that 52.0\% of the FN samples improved in relevancy score, and 12.9\% became relevant (TPs). These findings indicate the significance of FPE in recruitment-related RSs and suggest that further studies could help enhance the accuracy of these systems.

\begin{table}[]
    \centering
    \caption{\label{tab:fpe_adjectives}Examples of adjectives that exist in each unique set of job ads determined by Textblob~\cite{loria2018textblob}.}
    \begin{tabular}{l|l}
      Group  &  Adjectives \\
      \midrule
      Low Recall  & small, referral, sexual, steady, ... \\
      High Recall   & new, full, other, good, professional, ... \\
      Unique set   &  technical, annual, innovative, complex, ...\\
      \hline
    \end{tabular}
    \label{tab:unique_words}
\end{table}

\subsection{Ikea Effect}\label{sec:experiments:ikea}
\begin{figure} 
    \centering
    \scalebox{0.8}{
    \includegraphics[width=1\linewidth]{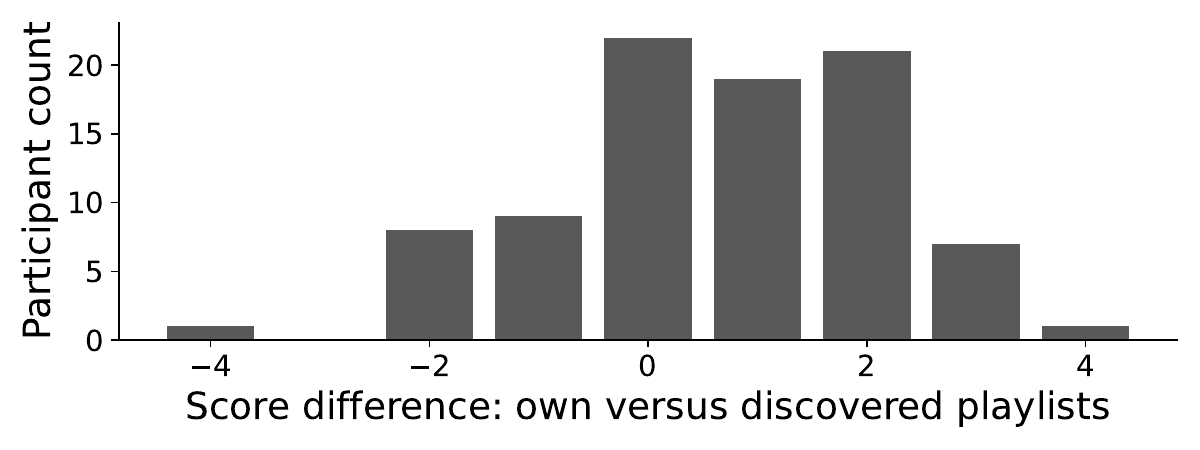}}\vspace{-4mm}
    \caption{Distribution of the consumption frequency difference between own and other playlists. Positive values show preference towards own playlists.}
    \label{fig:ikea_effect}
\end{figure}

In the context of RSs and streaming platforms, the Ikea Effect could be interpreted as a user's predisposition towards items and item collections they feel invested in (items they conducted research on or discovered themselves, collections they helped compile). To probe the effect in the music domain, we conduct a user study, striving to answer the following research question: Do music listeners prefer 
playlists they contributed to (\textit{own}) over playlists created without their participation (\textit{other})? We ran the study on the Prolific\footnote{\url{https://www.prolific.com}} platform for 100 participants from the United States who indicated themselves as users of one or more music streaming services. The study requires participants to complete four statements by choosing one option from a Likert-5 scale ranging from ``Never~(1)'' to ``Very often (5)''. The statements are presented in the same order to all participants: S1 ``I \textbf{create or edit} music collections...'',\footnote{The participants were instructed to interpret ``music collection'' as playlist, track compilation, mixtape or similar, created by a human or automatically.} S2 ``I \textbf{play} music collections (created by me or someone else).'', S3 ``I play music collections \textbf{I created or helped create myself}.'', S4 ``I play music collections \textbf{created by someone else} (e.g., discovered on the internet or shared by someone).''. 
Out of 96 respondents who submitted valid attempts, 88 indicated that they create or edit as well as consume playlists more often than "Never" (S1 and S2). Out of the 88 participants creating or editing playlists, 48 indicated that they consume own playlists more often than other (comparing responses to S3 and S4), and 18 participants indicated the opposite. The distribution of the difference in consumption frequency between own and other playlists (the S4 score subtracted from the S3 score for each user) is shown in Figure~\ref{fig:ikea_effect}. The corresponding mean of these (S3 score $-$ S4 score) is $0.65$ and the standard deviation is $1.52$. This shows that, in our sample, users tend to interact more with own playlists, which 
we interpret as an indicator of user preference and thus the Ikea effect. A larger study would be needed to evaluate the prevalence of the effect globally.
\vspace{-2mm}

\subsection{Cultural Homophily}\label{sec:experiments:homophily}
One of the possible ways to interpret the concept of homophily, i.e.,~the tendency of individuals to associate with similar others \cite{homophily2017}, in the context of RSs is through the interaction between item consumers and item creators. Users preferring items produced by creators sharing their cultural background would be an example of such interpretation~\cite{mark2003culture}.
In the following, we present the results of an empirical study showing indications of cultural homophily in the domain of music recommendation. We put forward two research questions: (1) Do users tend to prefer music tracks produced by artists from the same country? and (2) Can RSs foster or counteract homophily in the setting of music recommendation?
We answer the first question by analyzing the listening behavior of users from different countries. We answer the second question by conducting a feedback loop simulation, comparing recommendations provided by a RS at various steps of the feedback loop to user consumption behavior.

Following \cite{DBLP:conf/ismir/LesotaPBLRS22}, we select a 5-core filtered sample from years 2018-2019 of the LFM-2b dataset \cite{DBLP:conf/chiir/SchedlBLPPR22} containing 99,897 items (songs) selected uniformly at random and 2,287,732 interactions triggered by 11,776 users. We enrich the sample with information about the country of each artist crawled from MusicBrainz.\footnote{\url{https://www.musicbrainz.org}} We conduct a feedback loop simulation, following \cite{DBLP:conf/cikm/MansouryAPMB20}: At each of the 20 iterations, the data used for training on the previous step is enriched with one recommendation per user, selected from their personalized recommendations, with higher probability for higher ranking recommendations, and then used for training at the next iteration. We use MultVAE~\cite{Xu2021MultiVAE} as recommendation algorithm.

Table~\ref{tab:homophily} shows the proportions of domestic music in the data sample, user consumption behavior, and user recommendations for countries with the top 10 largest number of tracks in the data sample. The value $0.397$ for the $US$, $base$ shows that about $40\%$ of unique tracks in the data sample are produced by artists from the $US$. This column serves as a baseline, showing what the proportion of domestic consumption would be if every user were to interact with tracks chosen at random. The value $0.626$ for the $US$, $Con$ shows that over $60\%$ of actual listening attention of US users is directed towards tracks produced by US artists. Respectively $US$, $Rec_{20}$ of $0.595$ means that among all recommendations to US users at iteration $20$ under $60\%$ of tracks were produced by US artists.\footnote{Note that in the case of columns $Con$ and $Rec_*$ a track is counted multiple times if it is consumed by, or recommended to, multiple users (also taken into account for normalization).}

We answer the first question comparing the columns $base$ and $Con$, noticing that for all presented countries the proportion of actual consumption of domestic music is higher than the baseline. This shows that on average users have higher interest in music originating from their country than a random choice would warrant. We also notice that Finnish ($FI$) users demonstrate a higher level of interest in their domestic music than users from countries with comparable track supply, e.g., Australia ($AU$) and Brazil ($BR$).

Answering the second question we observe that, at the first iteration of the feedback loop ($Rec_1$), while users from the US get recommended a proportion of their domestic music similar to their actual consumption ($Con$), users from Germany, Brazil, and Russia get slightly over-served with their domestic music. Users from other countries are under-served with their domestic music, most prominently users from Australia, whose recommendations fail to reach the level of the baseline. At iteration 20 of the loop ($Rec_{20}$) the proportion of recommended domestic music increases for some under-served countries, such as the United Kingdom ($UK$) and Canada ($CA$). However, for other countries such as Sweden ($SE$) and Finland ($FI$) the proportion of recommended domestic music decreases further away from their actual consumption level.

The results of this preliminary study show that users tend to be interested in music originating from their country, potentially indicating cultural homophily. However, the degree of interest varies between countries. RSs do not necessarily represent this interest in their output and could foster it for some countries while counteracting it for others. 


\begin{table}
    \centering
    \caption{Proportions of domestic music among all available tracks ($base$), among consumed tracks by users from the country ($Con$), among recommended tracks ($Rec_1$), and among recommended tracks after 20 iterations of the feedback loop simulation ($Rec_{20}$), for the top 10 countries in the dataset. Bold values indicate the highest, underlined lowest values.\vspace{-3mm}} \label{tab:homophily}
    \scalebox{0.82}{
\begin{tabular}{lccccccc}
\toprule
 & $base$ & $Con$ & \cellcolor{gray!10} $Con/base$ & $Rec_1$ & \cellcolor{gray!10} $Rec_1/base$ & $Rec_{20}$ & \cellcolor{gray!10} $Rec_{20}/base$ \\
 \midrule
$US$ & $\textbf{0.397}$ & $\textbf{0.626}$ & \cellcolor{gray!10} $\underline{1.578}$ & $\textbf{0.629}$ & \cellcolor{gray!10} $1.587$ & $\textbf{0.595}$ & \cellcolor{gray!10} $1.501$ \\
$UK$ & $0.155$ & $0.266$ & \cellcolor{gray!10} $1.713$ & $0.227$ & \cellcolor{gray!10} $1.458$ & $0.232$ & \cellcolor{gray!10} $1.495$ \\
DE & $0.068$ & $0.169$ & \cellcolor{gray!10} $2.481$ & $0.176$ & \cellcolor{gray!10} $2.590$ & $0.166$ & \cellcolor{gray!10} $2.439$ \\
$SE$ & $0.045$ & $0.159$ & \cellcolor{gray!10} $3.519$ & $0.102$ & \cellcolor{gray!10} $2.266$ & $0.088$ & \cellcolor{gray!10} $1.948$ \\
$CA$ & $0.038$ & $0.083$ & \cellcolor{gray!10} $2.202$ & $0.030$ & \cellcolor{gray!10} $0.797$ & $0.041$ & \cellcolor{gray!10} $\underline{1.091}$ \\
$FR$ & $0.028$ & $0.091$ & \cellcolor{gray!10} $3.232$ & $0.039$ & \cellcolor{gray!10} $1.377$ & $0.041$ & \cellcolor{gray!10} $1.447$ \\
$AU$ & $0.023$ & $0.077$ & \cellcolor{gray!10} $3.289$ & $\underline{0.017}$ & \cellcolor{gray!10} $\underline{0.728}$ & $\underline{0.026}$ & \cellcolor{gray!10} $1.103$ \\
$FI$ & $0.023$ & $0.170$ & \cellcolor{gray!10} $\textbf{7.536}$ & $0.166$ & \cellcolor{gray!10} $7.325$ & $0.132$ & \cellcolor{gray!10} $5.820$ \\
$BR$ & $0.022$ & $0.141$ & \cellcolor{gray!10} $6.288$ & $0.187$ & \cellcolor{gray!10} $\textbf{8.347}$ & $0.150$ & \cellcolor{gray!10} $\textbf{6.714}$ \\
$RU$ & $\underline{0.019}$ & $\underline{0.073}$ & \cellcolor{gray!10} $3.870$ & $0.081$ & \cellcolor{gray!10} $4.262$ & $0.066$ & \cellcolor{gray!10} $3.515$ \\
\bottomrule
\end{tabular}}
\end{table}

\section{Conclusion}\label{sec:conclusions}
We demonstrated the presence and effects of several cognitive biases in the recommendation ecosystem, including feature-positive effect, Ikea effect, and cultural homophily. 
While the reported experiments are only first tiny steps, we hope to stimulate further research on this exciting topic.
Related to the biases studied here, future directions may include investigating the dependence of the Ikea effect on different functionalities offered by different streaming platforms and RSs which alters the ways users interact with them. Given the varying prevalence of certain RS platforms between countries, we expect notable differences.
The next steps in the study of cultural homophily could include disentangling homophily from other factors, such as availability bias, as well as studying the relation between homophily and diversity in recommendations.

We strongly believe that RSs researchers and practitioners should be equipped with high sensitivity for the existence of cognitive biases and knowledge of how they may impact different stages in the recommendation process.
We advocate for a holistic treatment of both the negative and positive effects of cognitive biases, also considering different stakeholders, which will be highly beneficial for the RSs community.  


\bibliographystyle{ACM-Reference-Format}
\balance
\bibliography{refs}





\end{document}